\documentclass[conference]{IEEEtran}
\IEEEoverridecommandlockouts

\usepackage{cite}
\usepackage{amsmath,amssymb,amsfonts}
\usepackage{algorithmic}
\usepackage{graphicx}
\usepackage{textcomp}
\usepackage{xcolor}
\def\BibTeX{{\rm B\kern-.05em{\sc i\kern-.025em b}\kern-.08em
    T\kern-.1667em\lower.7ex\hbox{E}\kern-.125emX}}

\usepackage{tikz}    
    
\usepackage{graphicx}
\usepackage[varg]{txfonts}

\usepackage{paralist}
\usepackage[abbrevmore]{bib_sty} 
\usepackage{subfigure}
\usepackage{color}
\usepackage{microtype}

\usepackage{url}

\newcommand{\Fig}{Fig.}
\newcommand{\Sec}{Sec.}

\newcommand{\dist}{v_j}

\newcommand{\nnrbP}{n_{s}}

\definecolor{lmag}{rgb}{0,0.82,1}

\definecolor{mag}{rgb}{0,0.7,.9}

\definecolor{magenta}{rgb}{1,0,1}
\definecolor{lblue}{rgb}{0.6,0.8,1}
\definecolor{lgreen}{rgb}{0,0.92,0.7}
\definecolor{lred}{rgb}{1,0.75,0.75}
\definecolor{lblue}{rgb}{0.6,0.8,1}
\definecolor{lgrey}{rgb}{0.85,0.85,0.85}

\definecolor{lye}{rgb}{1,0.9,0.6}

\newcommand{\Colorone}{solid blue}
\newcommand{\Colortwo}{dashed red}
\newcommand{\Colorthree}{dash-dotted green}
\newcommand{\Colorfour}{dotted magenta}

\newcommand{\GShat}{\widehat{GS}}

\newtheorem{procedure}{Procedure}
\newtheorem{remark}{Remark}
\newtheorem{thm}{Theorem}
\newtheorem{defn}{Definition}

\usepackage{eso-pic}

\begin{document}
\AddToShipoutPictureBG*{%
  \AtPageUpperLeft{%
    \setlength\unitlength{1in}%
    \hspace*{\dimexpr0.5\paperwidth\relax}
    \makebox(0,-0.75)[c]{ Tom Oomen, Learning for Advanced Motion Control, 
In  {\em IEEE International Workshop on Advanced Motion Control}, Agder, Norway, 2020}%
}}   

\title{Learning for Advanced Motion Control
}

\author{\IEEEauthorblockN{Tom Oomen}
\IEEEauthorblockA{
\textit{Eindhoven University of Technology, The Netherlands}\\
t.a.e.oomen@tue.nl}
}

\maketitle

\begin{abstract}
Iterative Learning Control (ILC) can achieve perfect tracking performance for mechatronic systems. The aim of this paper is to present an ILC design tutorial for industrial mechatronic systems. First, a preliminary analysis reveals the potential performance improvement of ILC prior to its actual implementation. Second, a frequency domain approach is presented, where fast learning is achieved through noncausal model inversion, and safe and robust learning is achieved by employing a contraction mapping theorem in conjunction with nonparametric frequency response functions. The approach is demonstrated on a desktop printer. Finally, a detailed analysis of industrial motion systems leads to several shortcomings that obstruct the widespread implementation of ILC algorithms. An overview of recently developed algorithms, including extensions using machine learning algorithms, is outlined that are aimed to facilitate broad industrial deployment.  

\end{abstract}


\section{Introduction}\label{sec:intro}

Learning from data has led to impressive achievements in recent years, many of which cannot go unnoticed in everyday life. Computer algorithms are now able to successfully learn in many domains, including human language such as speech recognition and accurate translations, real-time pattern recognition from images, digital advertising, self-driving vehicles, and in games such as Atari and Go \cite{SilberSchSimAntHuaGueHubBakLaiBolCheLilHuiSifDriGraHas2017}. The key enabler is the availability of large amounts of data in addition to ubiquitous and scalable computation and software. 

In sharp contrast, high-tech machines are often produced and installed with a pre-defined feedforward/feedback control algorithm, and their performance deteriorates over time due to wear, ageing, and varying environmental conditions such as temperature variations. Examples of such high-tech machines range from manufacturing machines such as lithographic wafer scanners \cite{Heertjes2016b}, \cite{BlankenBoeBruOom2017}, 2D and 3D printers, and pick-and-place robots, to scientific instruments such as microscopes \cite{AbramovitchAndPaoSch2007}, and medical equipment such as CT scanners. Interestingly, these high-tech machines are prime examples of mechatronic systems, where control algorithms are typically implemented in a digital computer environment. Hence, abundant data becomes available during the lifetime of these machines, yet this is often not exploited to enhance their performance.

Iterative Learning Control (ILC)~\cite{Moore1993}, \cite{BristowThaAll2006}, \cite{BienXu1998}, \cite{Gorinevsky2002} is a high-performance digital control strategy used to improve the performance of batch repetitive processes, by iteratively updating the command signal from one experiment to the next. Basically, ILC results in a command signal that can compensate for repeating components in the error signal, even if imperfect plant knowledge is available. ILC learns by updating the command input by filtering measured error data. To achieve fast learning, the filter should approximate the inverse of the closed-loop system. To achieve safe and robust learning, the approximation error should be sufficiently small. 

Many different ILC design frameworks have been developed and successful implementations have been reported. Design frameworks include frequency domain approaches \cite{SteinbuchMol2000}, optimization-based ILC \cite{TogaiYam1985}, \cite{GunnarssonNor2001}, Arimoto-type ILC \cite{ArimotoKawMiy1984}, and joint feedback and ILC design \cite{RogersGalOwe2007}, \cite{PaszkeRogGal2016}. Furthermore, theoretical aspects, e.g., convergence \cite{GhazaeiKhoBer2017}, are well-understood.

Although several ILC design frameworks are available, at present ILC is not yet broadly implemented as a standard industrial control component in advanced motion control of mechatronic systems. The aim of this paper is to provide a tutorial on ILC designs for advanced motion control in mechatronic systems, point out its shortcomings that obstruct widespread industrial deployment, and outline several recent developments that facilitate broad industrial deployment. These developments are described in detail in the references, and preliminary results of the case study appear in \cite{Oomen2018b}, \cite{BlankenZunRozStrOom2019}.

\section{From traditional motion control to learning}

\subsection{Motion control}
%

Precision mechatronics are essential for many industrial systems, including manufacturing machines and scientific instruments. Positioning systems are key subsystems that create the functionality in these machines. These subsystems are responsible for the motion that positions the product in the machine, e.g., the wafer in lithography, the substrate in printing, the sample in microscopy, and the mirror alignment in telescopes and lithographic optics. 

Positioning systems are mostly mechatronic systems that contain many aspects, including mechanics, electronics, sensors, actuators, and thermal conditioning, see \cite{MunnigschmidtSchEij2011}. A key aspect is motion control, where sensor measurements are processed by a digital controller to generate inputs for the actuators.  

\begin{figure}[t]
\centering
\includegraphics[width=.85\linewidth]{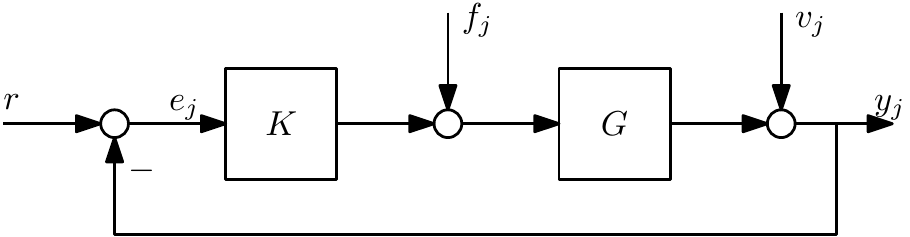}
\caption{Traditional motion control architecture.}
\label{fig:closedloop}
\end{figure}

Motion control often uses the architecture in \Fig~\ref{fig:closedloop}. Here, $G$ is the mechatronic system, including actuators, mechanics, and sensors. The goal is to track a reference $r$, i.e., minimize the error $e = r - y_j$. This is achieved through a feedback controller $K$  and feedforward signal $f_i$. In addition, $\dist$ is a disturbance.

\subsection{Reference design}\label{sec:reference}
\begin{figure}[t]
\centering
\includegraphics[width=.85\linewidth]{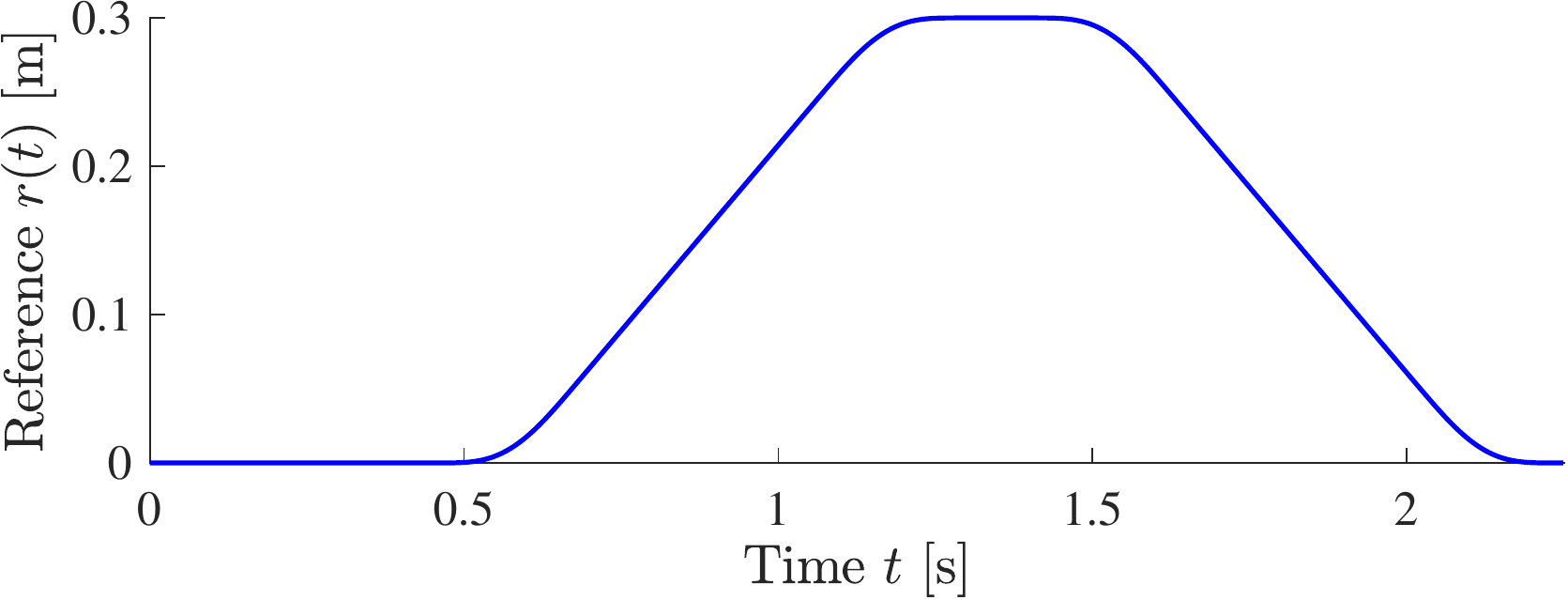}
\caption{Typical reference $r$.}
\label{fig:ILC5}
\end{figure}

Motion systems have to track a reference trajectory $r$. In certain applications, the goal is to perform a point-to-point motion, e.g., pick-and-place machines, including wire-bonders and die-bonders \cite{BoerenBarKokOom2016}, where the end-point accuracy is essential, see \cite{BoerenBruDijOom2014}. In other applications, e.g., printing systems and wafer scanners, the error should be small during constant velocity, see \cite{OomenRoj2017}. In both cases, $r$ is  designed as a $3^{\mathrm{rd}}$ or $4^{\mathrm{th}}$ order profile to ensure that actuator constraints are satisfied \cite{LambrechtsBoeSte2005}, see \Fig~\ref{fig:ILC5} for an example.

\subsection{Feedback control}\label{sec:feedback}
From \Fig~\ref{fig:closedloop}, the error in task $j$ is given by
\begin{equation}\label{eq:error}
e_j = S(r - Gf_j) - S \dist,
\end{equation}
where $S = \tfrac{1}{1+GK}$. In the case without feedforward, i.e., $f_j = 0$, then $e = S(r - \dist)$. A small error is then achieved by making $S$ small at the frequencies where the power spectrum of $r$ and $\dist$ is large. Typical references $r$ as described in \Sec~\ref{sec:reference} mostly have low-frequency content, hence $S$ must be made small at low frequencies. Similarly, $\dist$ has a certain power spectrum, and $S$ must be made small at those frequencies. Shaping these closed-loop functions such as $S$ while at the same time ensuring closed-loop stability is typically done using loop-shaping. Loop-shaping for motion systems typically leads to Proportional-Integral-Derivative (PID) type controllers, e.g., \cite{Oomen2020}. A typical constraint herein is the Bode sensitivity integral, which states that sensitivity reduction at low frequencies necessarily leads to an amplification at high frequencies, see \cite[\Fig\ 3 and 4]{Stein2003}. This relates to causality: the feedback controller is always too late, since it only takes action if the error is nonzero. 

\subsection{Feedforward control}\label{sec:FFcontrol}
The aim of feedforward is to compensate for reference-induced error signals before these affect the system. In view of \eqref{eq:error}, the goal of feedforward is to pick $f_j$ such that $r - Gf_j$ is minimized. This is achieved by selecting $f_j = G^{-1}r$. Determining the inverse of $G$ is typically done by bridging first-principles system knowledge with data-driven tuning. In particular, typical motion systems are of the form \cite{Gawronski2004} 
\begin{equation}\label{eq:modal}
G =\underbrace{\frac{1}{ms^2}}_{\mathrm{rigid-body\ mode}} + \underbrace{\sum_{i=N_{rb}+1}^{\nnrbP}\frac{r_i}{s^2+2\zeta_i\omega_is+\omega_i^2}}_{\mathrm{flexible\ modes}},
\end{equation}
hence in the low-frequency range, where $r$ has most of its energy, the system is approximately a double integrator, i.e., $\tfrac{1}{ms^2}$. As a consequence, $G^{-1}$ is approximated as $ms^2$, leading to $f_j = m s^2 r$. Since the inverse Laplace transform of $s^2r$ is given by the acceleration profile $\ddot r$, $f_j = m \ddot r$. The parameter $m$ is tuned using measured data, see \cite{Oomen2020}, \cite{Heertjes2016b}. Note that \eqref{eq:modal} also gives rise to other feedforward parameters such as snap \cite{Oomen2020}. 
 
\subsection{Learning from data and requirements}

Both feedback and feedforward control are used as standard components for motion control of precision mechatronics. However, in practice always an error remains, which is measured directly in the system. Indeed, data is abundantly available in present-day mechatronic systems, yet this information is still far from fully exploited. Indeed, sensors in mechatronic systems are often used for feedback control, which typically only make use of real-time position and velocity information in PID-type controllers. The aim here is to fully exploit all data from past and already completed tasks to achieve control to the limit of the predictable behavior of mechatronic systems.

Learning in mechatronic systems imposes several unique requirements, since it involves interactions with the real world. The following requirements are considered throughout.
\begin{enumerate}
\item Learning should be fast, since machines require experiments in real-time. In addition, fast adaptation is useful in case of varying conditions, e.g., time variations over different periods, e.g., day/night and seasonal changes. 
\item Learning should be safe and use operational data, since dedicated experiments may induce production loss and even damage of the machine. 
\end{enumerate}


This paper addresses three questions. First, what can learning achieve for a specific system? Second, how to achieve fast and safe learning convergence using operational data? Third, why is learning not ubiquitous in industrial motion control? These questions are addressed in the forthcoming sections. 

\begin{remark}
Throughout, the system resets after each task, leading to iterative learning control with both a time and task variable, see \Fig~\ref{fig:ILC1}. In case of continuous operation, repetitive control is obtained, for which a similar design framework is available yet different analysis tools are required, see \cite{Longman2000}, \cite{BlankenKoeOom2020}.\end{remark}
%

\section{What does ILC have to offer for a specific system?}\label{sec:ILCoffer}

\begin{figure}[t]
\centering
\includegraphics[width=.75\linewidth,page=1]{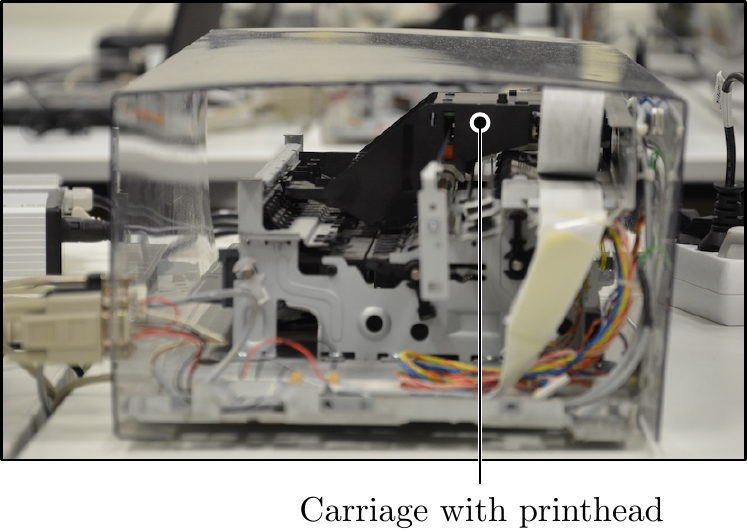}
\caption{Printer system with repeating tasks.}
\label{fig:printer}
\end{figure}

\begin{figure}[t]
\centering
\includegraphics[width=.85\linewidth,page=1]{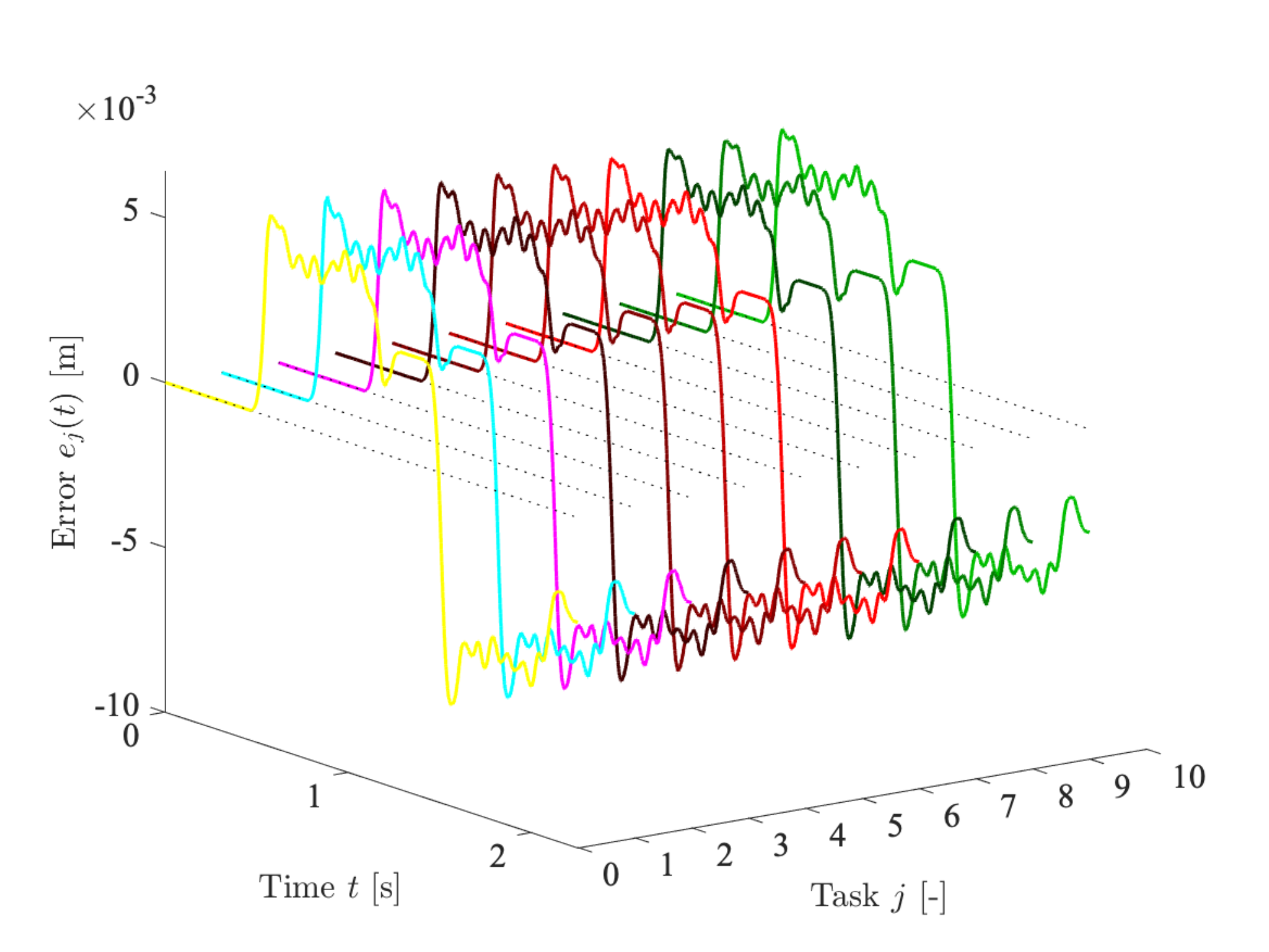}
\caption{Measured error signal for ten tasks $j$ using only feedback control.}
\label{fig:ILC1}
\end{figure}

\begin{figure}[t]
\centering
\includegraphics[width=.85\linewidth,page=1]{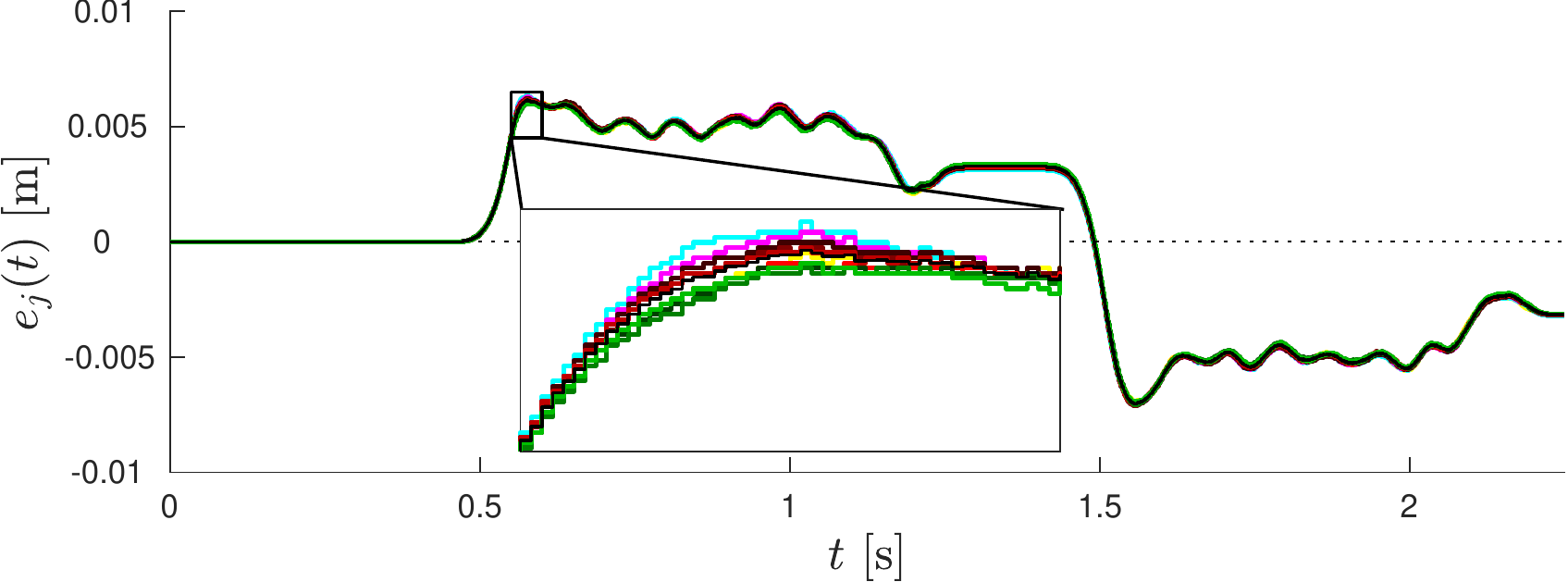}
\caption{Measured error signal for ten tasks $j$ using only feedback control of \Fig~\ref{fig:ILC1}, together with sample mean $m_e(t)$ (solid black). The main idea is that the reproducible part $m_e(t)$ of the error can be easily predicted and hence compensated.}
\label{fig:ILC1b}
\end{figure}

\begin{figure}[t]
\centering
\includegraphics[width=.85\linewidth,page=1]{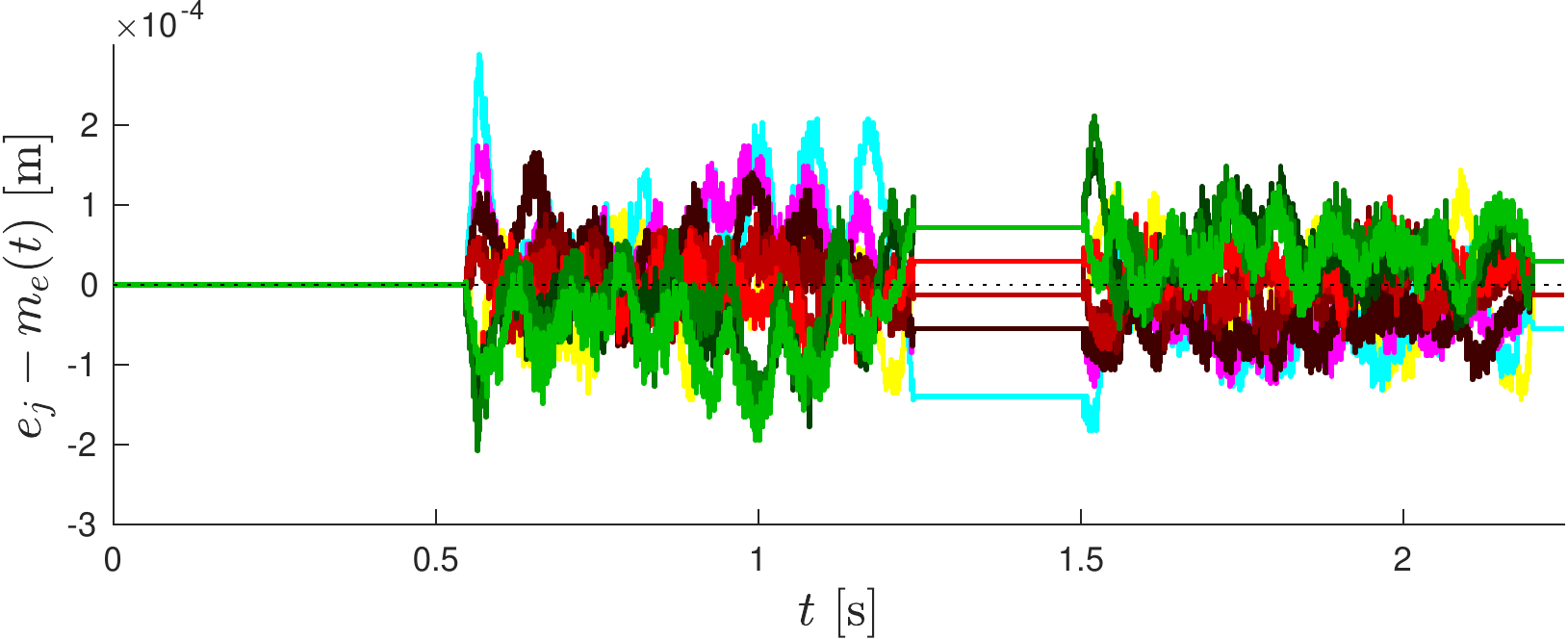}
\caption{Measured error signal for ten tasks $j$ using only feedback control of \Fig~\ref{fig:ILC1}, where the sample mean $m_e$ of \Fig~\ref{fig:ILC1b} has been subtracted. The remaining non-reproducible part of the error remains, which typically cannot be compensated using ILC. This is the order of magnitude of the error that ILC will be able to obtain, which is substantially less than the error using feedback only. }
\label{fig:ILC1c}
\end{figure}

A main advantage of ILC is that the achievable performance can be estimated before actually implementing ILC, providing the user with an insightful performance lower bound. This is illustrated using experiments on the desktop printer in  \Fig~\ref{fig:printer}. The goal is to position the carriage that contains the printheads. The input of the system is the voltage signal to the motor that drives the carriage through a rubber belt. The output is the position of the carriage, which is measured using a linear encoder. Throughout, a PD feedback controller is implemented. 

\newcommand{\nexp}{n_\mathrm{exp}}

The main idea is that ILC can compensate the reproducible and hence predictable part of the error. To determine this part, ten identical tasks have been performed using only feedback control, corresponding to printing ten lines on a sheet of paper. The resulting error signals are depicted in \Fig~\ref{fig:ILC1}. Clearly, the error is highly reproducible. To further analyze the error signal, let the error in task $j$ be denoted $e_j(t)$, $j = 0, \ldots, \nexp -1$, where $\nexp = 10$ denotes the number of tasks. 

Next, compute the sample mean of the error signal 
\begin{equation}
m_e(t) = \frac{1}{\nexp} \sum_{j=0}^{\nexp-1} e_j(t).
\end{equation}
For the repeated tasks in \Fig~\ref{fig:ILC1}, the sample mean $m_e$ is depicted in \Fig~\ref{fig:ILC1b}, together with the ten realizations $j$. 

Clearly, the part $m_e$ is easy to predict and should be possible to compensate by designing $f_j$ in \Fig~\ref{fig:closedloop}. By subtracting the sample mean $m_e$ from the realizations $e_j(t)$, the non-reproducible part remains, see \Fig~\ref{fig:ILC1c}. This is the residual error that cannot be compensated directly using ILC. Typically, this residual error due to non-reproducible disturbances is at least an order of magnitude smaller compared to $m_e$, see \Fig~\ref{fig:ILC1c}. 

Summarizing, an estimate of the potential performance increase of ILC is obtained, i.e., reducing the error from \Fig~\ref{fig:ILC1b} to \Fig~\ref{fig:ILC1c}. The remaining question is how to actually achieve this, which is investigated in the forthcoming section. 


\begin{remark}
In Fig~\ref{fig:ILC1}-\ref{fig:ILC1c}, an analysis has been done using feedback only. A similar approach can be pursued in case feedforward is already implemented. 
\end{remark}

\begin{remark}
The reproducible part, i.e., the sample mean $m_e$ in \Fig~\ref{fig:ILC1b}, can be directly compensated using ILC. Clearly, the residuals contain non-repeating parts that can be mitigated using feedback, see \cite{ZundertOom2017e} for details. 
\end{remark}

\section{Frequency domain ILC for precision mechatronics} \label{sec:freqdomILC}

\subsection{A basic approach for fast learning from past data}\label{sec:basicILC}

\begin{figure}[t]
\centering
\includegraphics[width=.65\linewidth]{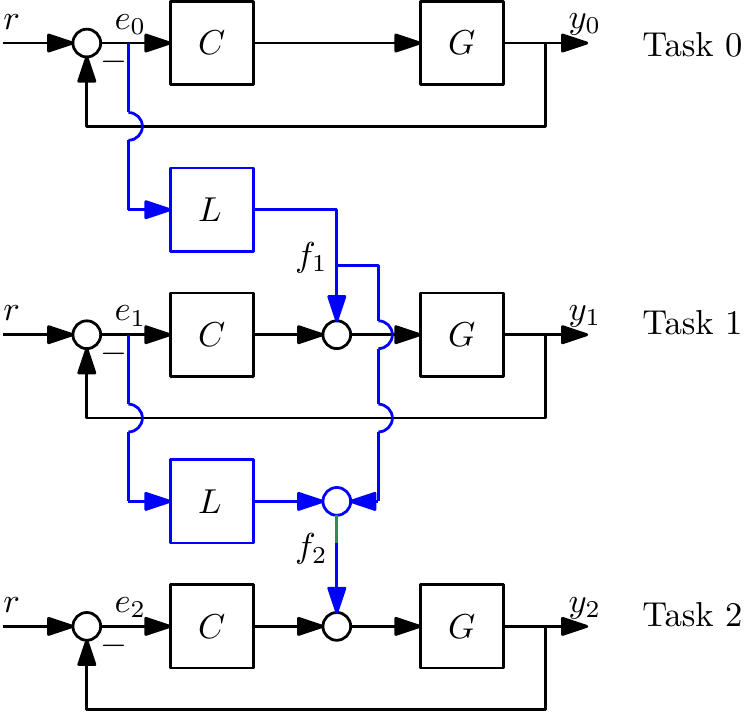}
\caption{Learning from past tasks.}
\label{fig:Figure5b}
\end{figure}

The main idea in ILC is to learn from past measured error signals. Suppose that an initial task has been completed. Consider for example task $j = 0$ in \Fig~\ref{fig:ILC1}, where a feedback controller is used and feedforward $f_0 = 0$. 

After each task $j$, the ILC algorithm should generate a command signal for the next task $f_{j+1}$. Hence, command signal $f_{1}$ is determined prior to starting task $1$, i.e., based on measured data in task $0$. Assume for the moment that the same task is performed, and that only access to measured error signal $e_0$ is available. 
In addition, the disturbance $\dist$ is zero. 

This leads to the following problem: determine $f_1$ based on the measured signal $e_0$, such that $e_1$ is small. Note from \Fig~\ref{fig:Figure5b} that 
\begin{align}
e_0 &= S r \label{eq:e0}\\
e_1 &= Sr - GS f_1.\label{eq:e1}
\end{align}
Next, the first key step is to substitute \eqref{eq:e0} into \eqref{eq:e1}, leading to
\begin{equation}
e_1 = e_0 - GS f_1.
\end{equation}
The second key step is to pick $f_1$ as a filtered version of $e_0$, see also \Fig~\ref{fig:Figure5b}, i.e., according to the update law
\begin{equation}
f_{1} = L e_0.
\end{equation}
In this case, 
\begin{equation}
e_1 = (1 - GSL)e_0.
\end{equation}
Next, $e_1 = 0$ is obtained by $(1-GSL) = 0$. This is achieved by $L = (\GShat)^{-1}$, where $\GShat$ is a model of the true system $G$. 

In practice, model errors, i.e., $\GShat \neq GS$ may lead to a situation where $f_1$ leads to an error signal $e_1$ that is not exactly zero. This leads to the concept of iterative learning control:
\begin{equation}\label{eq:learningnorobustness}
f_{j+1} = f_j + L e_j.
\end{equation}
The intuition is as follows: if $f_1$ already leads to $e_1 = 0$, then in the next task $j = 2$ this command input is retained, i.e., $f_2 = f_1$. Otherwise, a small correction $Le_1$ is added to $f_1$. 

\subsection{The need for robustness for safe learning} \label{eq:ILCisfeedback}

\subsubsection{Implementation of the learning update}\label{sec:divergence}
\begin{figure}[t]
\centering
\includegraphics[width=.85\linewidth,page=1]{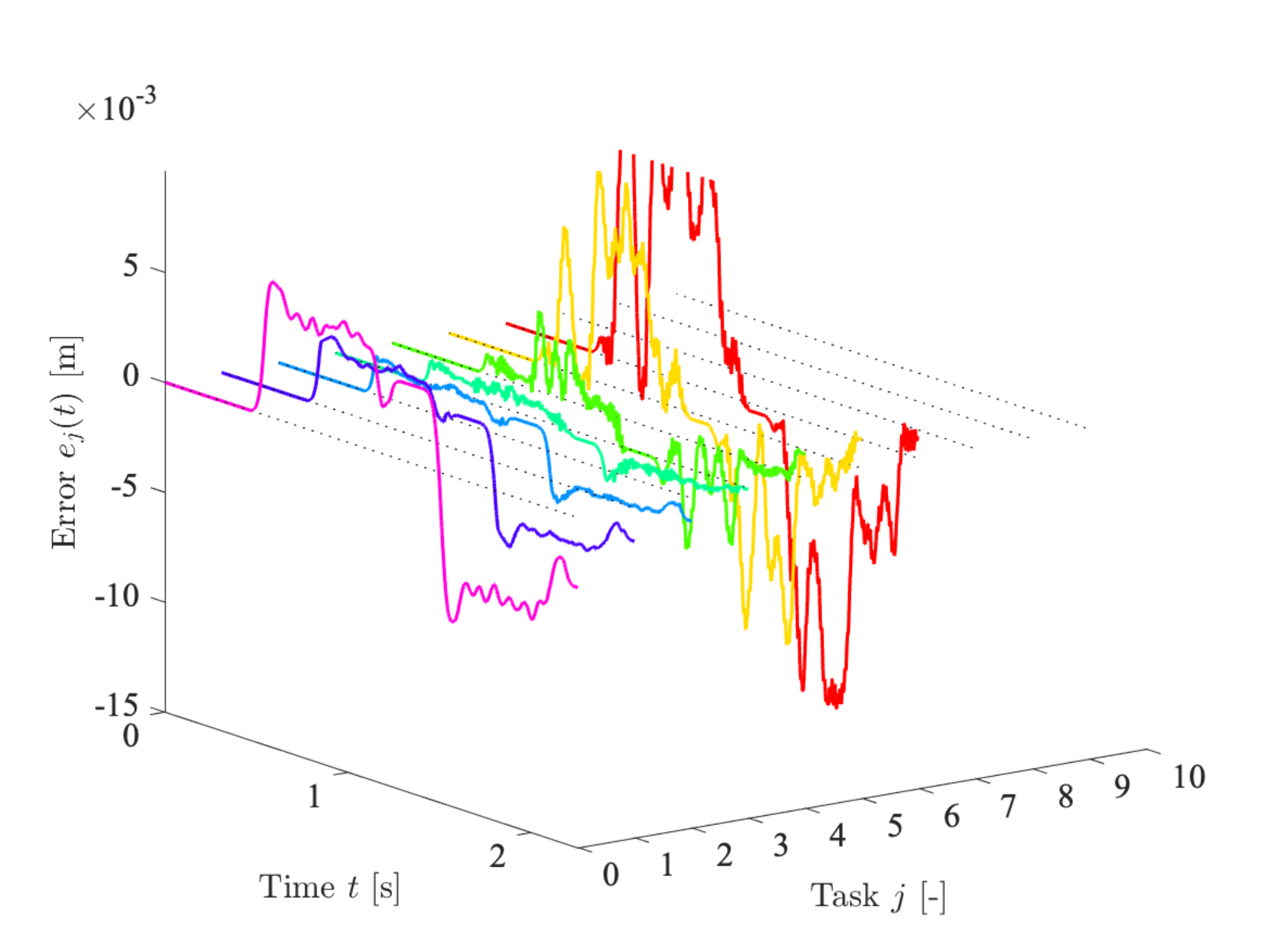}
\caption{Measured error signal for ten tasks $j$ using the learning update \eqref{eq:learningnorobustness}. In the first three tasks, the error signal reduces, but after four tasks it starts to increase.}
\label{fig:ILC3}
\end{figure}

\begin{figure}[t]
\centering
\includegraphics[width=.85\linewidth,page=1]{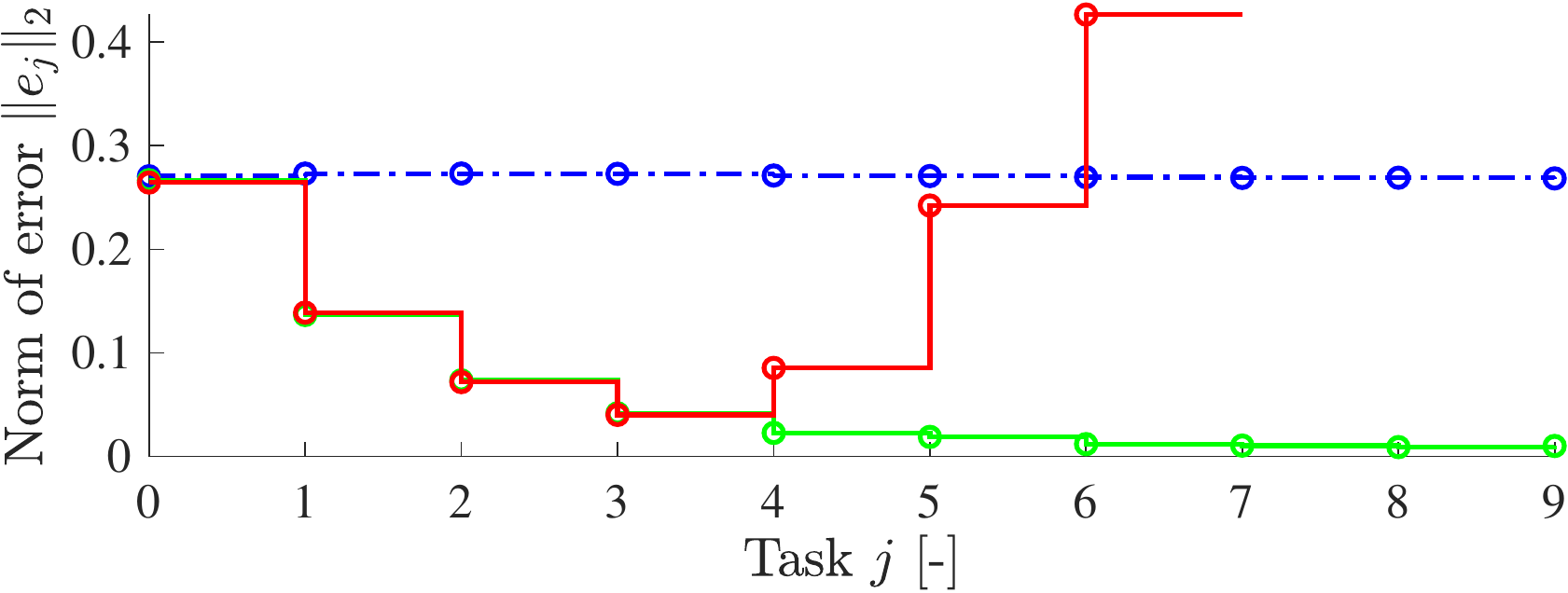}
\caption{Measured error signal for ten tasks $j$ using feedback only (blue), the learning update \eqref{eq:learningnorobustness} (red), learning update \eqref{eq:learningQ} (green).}
\label{fig:ILC4}
\end{figure}

Implementation of the learning algorithm~\eqref{eq:learningnorobustness} during ten tasks as in \eqref{fig:ILC1} leads to the error signal in \Fig~\ref{fig:ILC3}. In the first tasks, i.e., $j = 1,2,3$, the error reduces. However, from task $4$ onwards, the error starts to increase again. This is also clearly visible from \Fig~\ref{fig:ILC4}, where the $2$-norm of the energy is depicted. This is an undesired aspect, which needs further investigation.

\subsubsection{Convergence analysis}
The example in \Sec~\ref{sec:divergence} reveals that learning can lead to unbounded error signals. The main reason is that learning generates a feedforward signal in each task, see \Fig~\ref{fig:Figure5b}, yet leads to a feedback in the iteration domain. In particular, to analyze convergence, note from \eqref{eq:e1} that 
\begin{equation}\label{eq:ej}
e_j = Sr - GSf_j.
\end{equation}
Also, consider the extended learning update
\begin{equation}\label{eq:learningQ}
f_{j+1} = Q( f_j + L e_j),
\end{equation}
where \eqref{eq:learningnorobustness} is recovered by setting $Q=1$. By evaluating \eqref{eq:ej} for both $j$ and $j+1$, and using \eqref{eq:learningQ}, this leads to
\begin{equation}\label{eq:iteration}
e_{j+1} = Q(1-GSL)e_j+(1-Q)Sr.
\end{equation}
The key question is whether the iteration \eqref{eq:iteration} converges. The following theorem gives a decisive answer to this issue through a contraction mapping. Here, the notion of monotonic convergence is used, which is defined as follows. 
\begin{defn}
The iteration \eqref{eq:iteration} is monotonically convergent in the $\ell_2$ norm if, for some $k \in [0,1)$ and for all $e_j$,
\begin{equation}
\|e_\infty - e_{j+1}\|_2 \leq k \|e_\infty - e_j\|_2.
\end{equation}
\end{defn}

\begin{thm}\label{thm:monconv}
The iteration \eqref{eq:iteration} is monotonically convergent in the $\ell_2$ norm to a fixed point $e_\infty$ and corresponding $f_\infty$ if and only if 
\begin{equation}\label{eq:convergencecondition}
\left \| Q(1-GSL)\right\|_{\mathcal{L}_\infty} < 1.
\end{equation}
\end{thm}
For a proof of Theorem~\ref{thm:monconv}, see \cite[Theorem 2]{OomenRoj2017}. Here, the $\ell_2$ norm of a signal $x$ is defined as $\|x\|_2 = \sqrt{\sum_{t=-\infty}^{\infty} |x(t)|^2}$. Also, the $\mathcal{L}_\infty$ norm for a single-input single-output system $F$ is defined as $\|F\|_{\mathcal{L}_\infty} = \sup_\omega \left | F (e^{j\omega })\right |$, hence \eqref{eq:convergencecondition} is equivalent to the frequency domain test
\begin{equation}\label{eq:condition}
\left | Q(1-GSL) \right|<1 \forall \omega.
\end{equation}
Note the resemblance of the $\mathcal{L}_\infty$ norm with the commonly used $\mathcal{H}_\infty$ norm. The key difference is that the $\mathcal{L}_\infty$ allows for non-causal ILC algorithms, i.e, non-causal $L$ and $Q$ in \eqref{eq:learningQ}. This is essential for ILC and does not require $L$ and $Q$ to be analytic outside the unit disc, as is explained in \Sec~\ref{sec:implementation}.

The most important aspect for practical mechatronic systems is the fact that frequency response function measurements are fast, accurate, and inexpensive \cite{MaasMaasVooOom2017}. Interestingly, the choice $L = (GS)^{-1}$  in \Sec~\ref{sec:basicILC} typically involves a parametric model, but condition \eqref{eq:condition} can be directly verified for a nonparametric frequency response function, possibly for a range of relevant operating conditions \cite{MaasMaasVooOom2017} or for a range of systems to address machine-to-machine variability. 


Application of Theorem~\ref{thm:monconv} to analyse the situation of \Fig~\ref{fig:ILC3} leads to the result in \Fig~\ref{fig:ILC11}, where an identified frequency response function of the printer is used. Clearly, condition \eqref{eq:condition} is violated, hence the iteration \eqref{eq:iteration} does not converge.

\newcommand{\Lacc}{L^{\mathrm{accurate}}}
\newcommand{\Qacc}{Q^{\mathrm{accurate}}}

\begin{figure}[t]
\centering
\includegraphics[width=.85\linewidth,page=1]{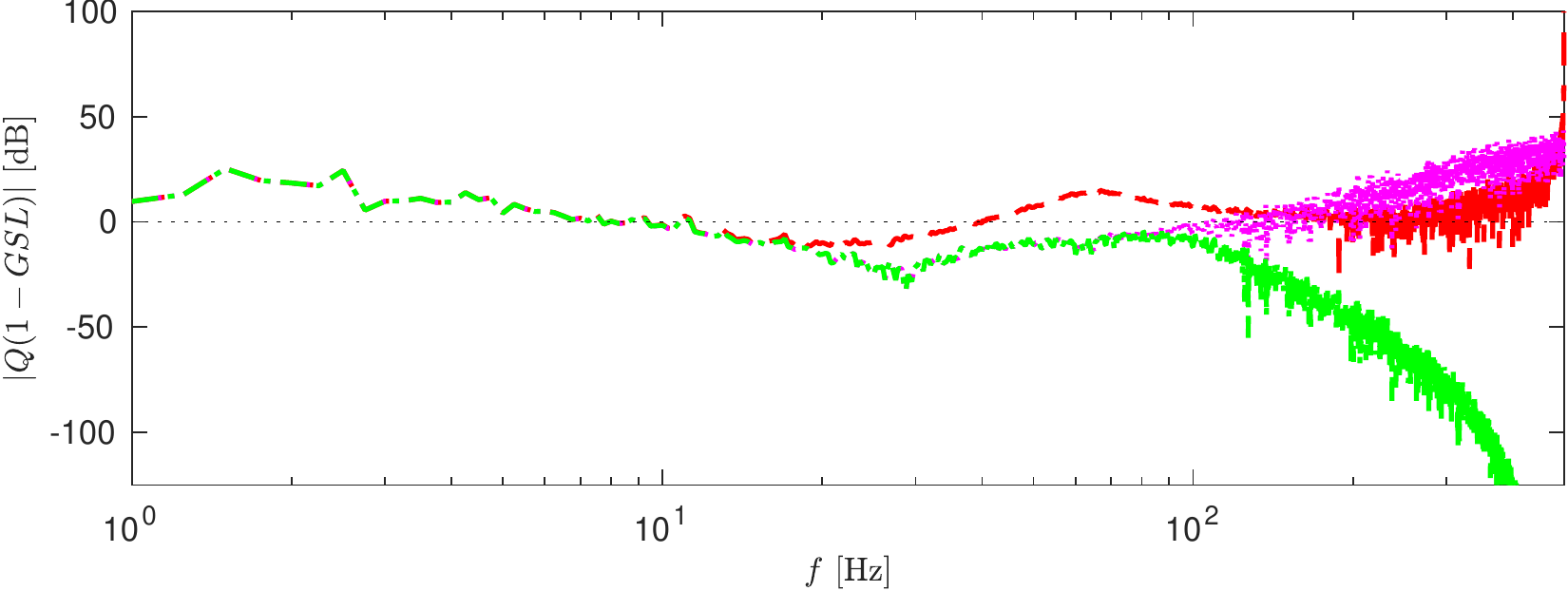}
\caption{Convergence condition~\eqref{eq:condition} for the ILC iteration of \Fig~\ref{fig:ILC3} (\Colortwo). Next, \eqref{eq:condition} is verified for the new model, i.e., $1-GS\Lacc$ (\Colorfour). Finally, robust monotonic convergence is guaranteed by inclusion of $Q$, leading to $\Qacc(1-GS\Lacc)$ (\Colorthree). In the latter case, the convergence condition is met for frequencies beyond $10 \ \mathrm{Hz}$. Below $10 \ \mathrm{Hz}$, the frequency response function is very noisy, and therefore discarded.}
\label{fig:ILC11}
\end{figure}

\begin{figure}[t]
\centering
\includegraphics[width=.85\linewidth,page=1]{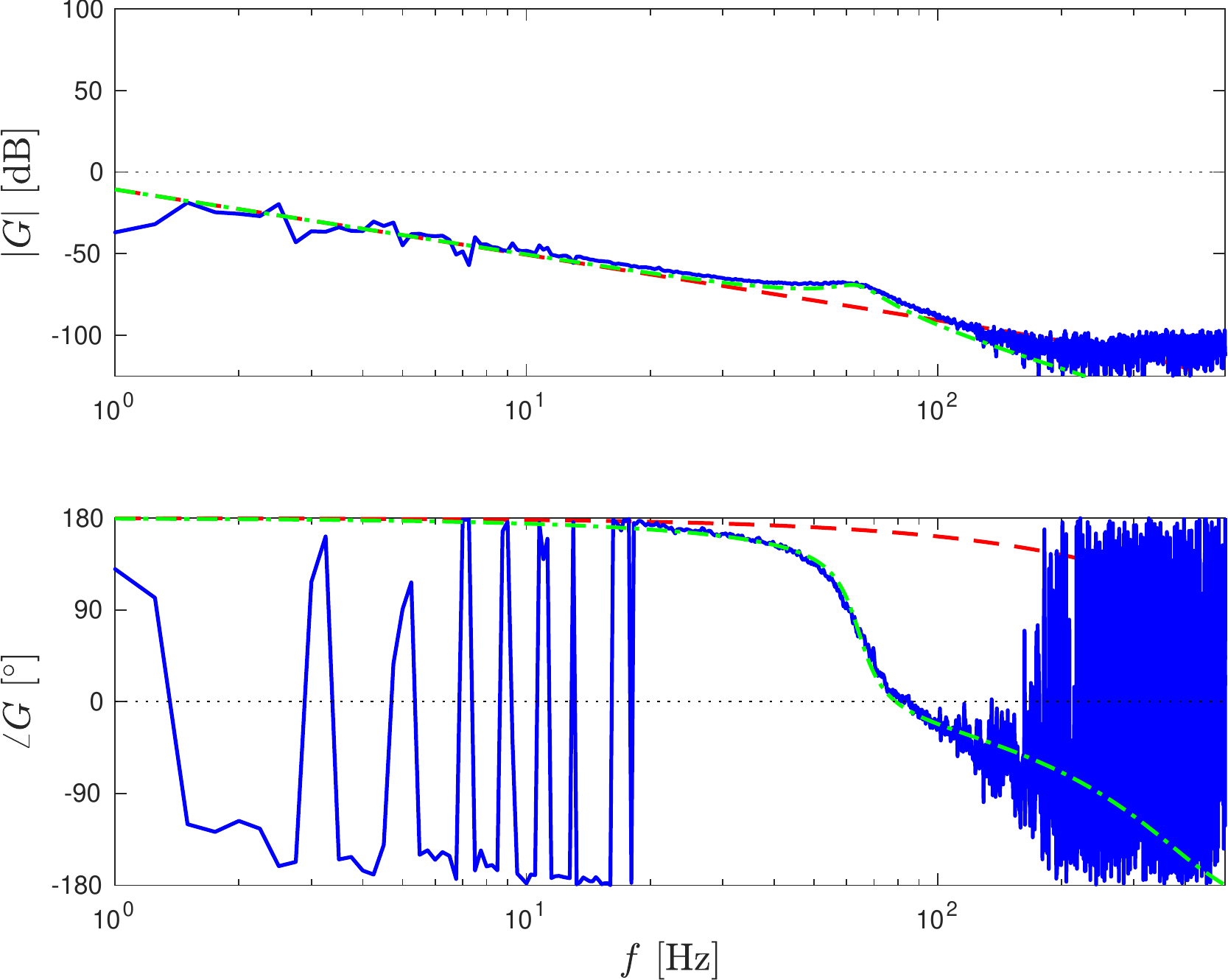}
\caption{Identified frequency response function of the printer system (\Colorone), parametric model corresponding to \Fig~\ref{fig:ILC3} (\Colortwo), and improved parametric model corresponding to \Fig~\ref{fig:ILC2} (\Colorthree).}
\label{fig:ILC10}
\end{figure}

\subsubsection{Designing ILC for robust monotonic convergence}\label{sec:ILCdesign}

To design a robust monotonically convergent ILC, the following design procedure is considered. 
\begin{compactenum}
\item Determine a parametric model $\widehat{GS}$ of $GS$ and determine $L = (\widehat{GS})^{-1}$.\label{step:performance}
\item Determine $Q$ such that $Q = 1$ for frequencies where $1-GSL <1$, and $|Q|<1$ such that $|Q(1-GSL)|<1, \forall \omega$.\label{step:robustness}
\end{compactenum}

In view of the printer setup, i.e., \Fig~\ref{fig:ILC11}, both $L$ and $Q$ are designed. To address Step~\ref{step:performance}, the model of the system is improved, see \Fig~\ref{fig:ILC10}. In particular, the model corresponding to \Fig~\ref{fig:ILC3} only contains the rigid-body mode-shape in \eqref{eq:modal}. The model is extended with a flexible mode, corresponding to the flexibility introduced by the belt mechanism, see \Fig~\ref{fig:printer}. This leads to $\Lacc$, which already improves the convergence condition, see \Fig~\ref{fig:ILC11}. Next, to guarantee convergence, $\Qacc$ is designed in Step~\ref{step:robustness}, see \Fig~\ref{fig:ILC11}. Note that $Q$ should be chosen close to $1$ to ensure high performance. Indeed, $Q \neq 1$ leads to an asymptotic error, which can be directly derived from \eqref{eq:iteration} and \eqref{eq:e0} and $e_\infty = \lim_{j\rightarrow \infty}e_j$, leading to $e_\infty = \frac{1-Q}{1-Q(1-GSL)}e_0$.

\begin{figure}[t]
\centering
\includegraphics[width=.85\linewidth,page=1]{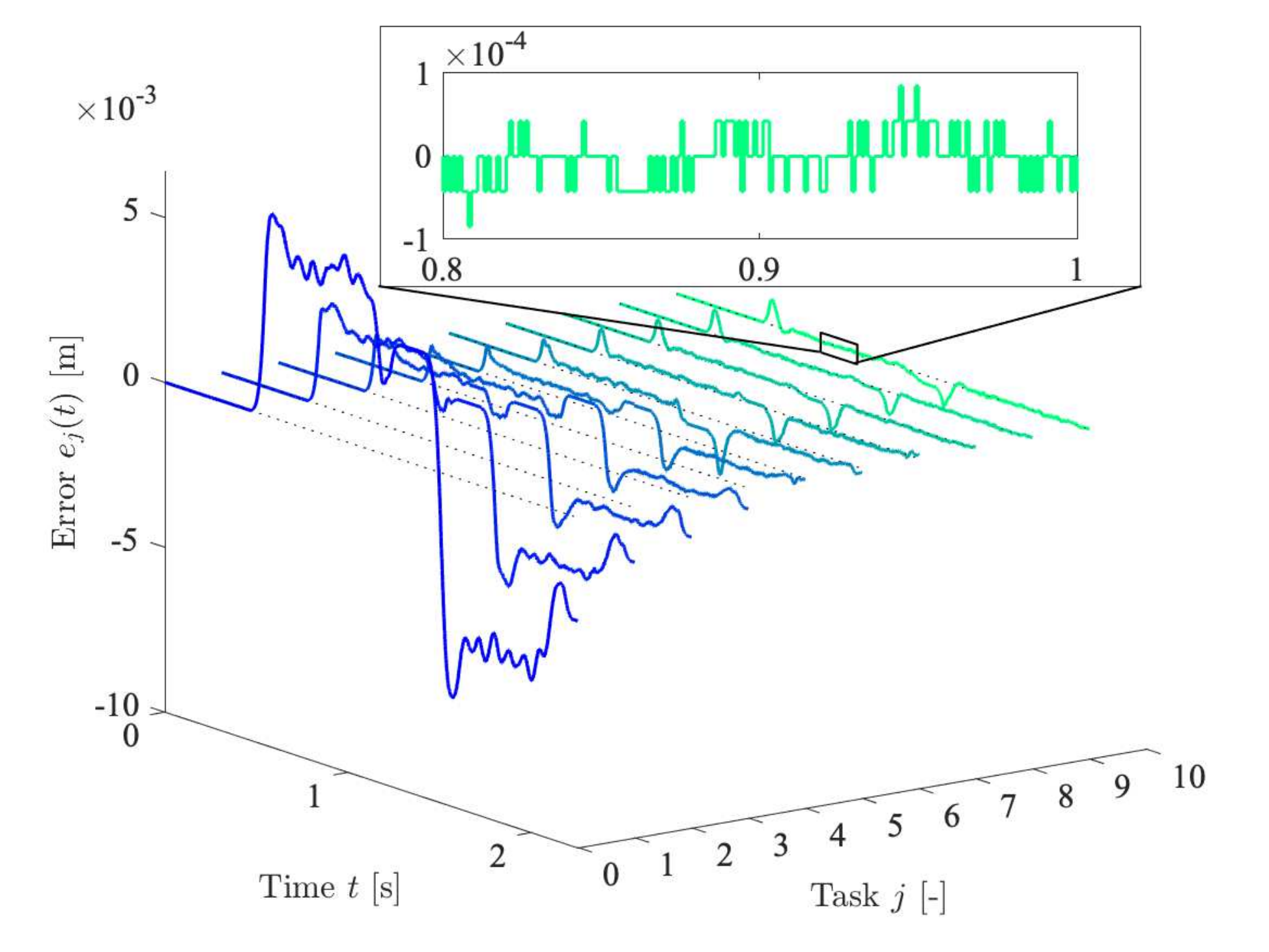}
\caption{Measured error signal for ten tasks $j$ using the learning update \eqref{eq:learningQ} using $\Lacc$ and $\Qacc$ of \Fig~\ref{fig:ILC11}. The error reduces monotonically towards $e_\infty$ for increasing $j$, achieving almost encoder resolution.}
\label{fig:ILC2}
\end{figure}

The resulting $\Lacc$ and $\Qacc$ are implemented on the system, see \Fig~\ref{fig:ILC2} for results. Clearly, the error is reduced substantially towards the encoder resolution. Hence, after $10$ iterations it is already in line with \Sec~\ref{sec:ILCoffer}.

\begin{remark}
Many alternative ILC designs are available, e.g., Arimoto ILC, where $L$ is designed as a proportional-derivative filter. A key advantage of the approach outlined here is its fast convergence in conjunction with a systematic design of robustness filters using non-parametric frequency response functions, which is particularly well suited for mechatronic systems. Note that the convergence results, including Theorem~\ref{thm:monconv}, can be directly applied to alternative ILC designs. 
\end{remark}

\subsubsection{Implementation aspects}\label{sec:implementation}
\begin{compactenum}
\item One of the key benefits of ILC is the fact that it can generate non-causal signals in the time domain by appropriate design of $L$. Indeed, in case the system has a relative degree of at leat two, feedback control is subject to the Bode sensitivity integral, see \Sec~\ref{sec:feedback}. An example is the case where $GS$ has strict delay, e.g., $GS = z^{-d}$. In this case, the optimal $L$ is given by $L = z^d$, i.e., $d$ samples preview. To see this, note that $e_j = Sr - z^{-d} f_j$, hence $f_j$ is always delayed by $d$ samples, which must be compensated by $L$, i.e., there must be an information flow to earlier time instants. Similarly, in many cases $GS$ has non-minimum phase zeros, hence inversion may lead to unstable poles. Both cases motivate a non-causal design of $L$. For ILC, stable inversion techniques enable infinite preview, while $\mathcal{H}_\infty$ preview control provides an optimal solution for the finite preview case \cite{ZundertOom2018c}.
\item Similarly, $Q$ is designed in Step~\ref{step:robustness} of \Sec~\ref{sec:ILCdesign} such that $|Q(1-GSL)|<1 \forall \omega$. Note that the phase of $Q$ is does not affect robustness, but negatively influences performance. Therefore, the phase is eliminated by filtering with a filter $\tilde Q$ and its adjoint $\tilde Q^*$  \cite{BolderKleOom2018}. In this case, the convergence condition becomes $|\tilde Q ^* \tilde Q(1-GSL)| = |\tilde Q|^2|(1-GSL)|$. 
\item ILC may lead to an amplification of measurement noise, which can be mitigated by a learning gain \cite[\Sec\ 3]{OomenRoj2017}.
\item In case the feedback controller $K$ in \Fig~\ref{fig:closedloop} contains an integrator, then the inversion step in Step~\ref{step:performance} in \Sec~\ref{sec:ILCdesign} is troublesome. See \cite[\Sec\ 6]{BolderOom2016} for a solution.
\end{compactenum}

%
%
%

\section{Towards industrial use in complex mechatronics} \label{sec:wideinindustry}

The systematic ILC design framework for mechatronic systems in \Sec~\ref{sec:freqdomILC} allows for a substantial performance improvement, still its full potential in industrial application is largely unexploited. The aim of this section is first to investigate industrial requirements for ILC algorithms. This reveals key reasons that have led to limited industrial adoption. In addition, an overview of recent developments that aim to facilitate their industrial deployment is provided.

\subsection{Automated feedforward tuning for flexible tasks}\label{sec:flextasks}

Learning control can potentially compensate for all repeating disturbances, i.e., iteration-invariant disturbances that are identical for each task. Indeed, in \Sec~\ref{sec:freqdomILC}, it has only been assumed that $r$ is constant, it need not be known or directly measurable. However, in many mechatronic applications, including printing systems \cite{BolderOomKoeSte2014c}, wafer scanners \cite{BlankenBoeBruOom2017}, semiconductor backend equipment \cite{BoerenBarKokOom2016}, and additive manufacturing \cite{GuoPetOomMis2018}, setpoints change each iteration. Small variations can already lead to a disastrous effect. To see this, let $r_j$ depend on the task. Following the approach in \Sec~\ref{sec:basicILC}, this leads to $e_1 = Sr_1 - GSL e_0$, where $e_0 = Sr_0$, hence if $L = (GS)^{-1}$ then $e_1 = S(r_1 - r_0)$.  

To cope with setpoint variations, several frameworks have been developed. These include approaches where the setpoint is built up from a library of subtasks \cite{HoelzleAllWag2011}, initialization based on model knowledge \cite{JanssensPipSwe2012}, use of Artificial Neural Networks (ANNs) \cite{Moore1993}, and basis function ILC. Regarding the latter, this includes the use of polynomial basis functions, which resembles traditional feedforward where the parameters such as $m$ as described in \Sec~\ref{sec:FFcontrol}  are automatically learned  \cite{WijdevenBos2010} \cite{MeulenTouBos2008}, see also \cite{FruehPha2000} for an early result in this direction. Further developments regarding input shaping are presented in \cite{BoerenBruDijOom2014} and rational systems in \cite{BolderOom2015}, \cite{BlankenBoeBruOom2017}. These approaches typically employ the so-called lifted formulation, involving finite time signals and optimization criteria as in \cite{GunnarssonNor2001}. In \cite{BoerenBarKokOom2016}, basis functions are incorporated in the design approach of \Sec~\ref{sec:freqdomILC}.

Finally, an important aspect in basis function ILC is the actual selection of basis functions. This is essentially a model order selection problem, which connects to recent results in machine learning. It is addressed in \cite{OomenRoj2017} from a sparse optimization viewpoint. A fundamentally different approach is taken in \cite{BlankenOom2020} by using kernel-based regression, essentially viewing the system as a Gaussian Process (GP), see also \cite{PillonettoDinCheDenLju2014}. 

\begin{figure}[t]
\centering
\fbox{\includegraphics[width=.75\linewidth,page=1]{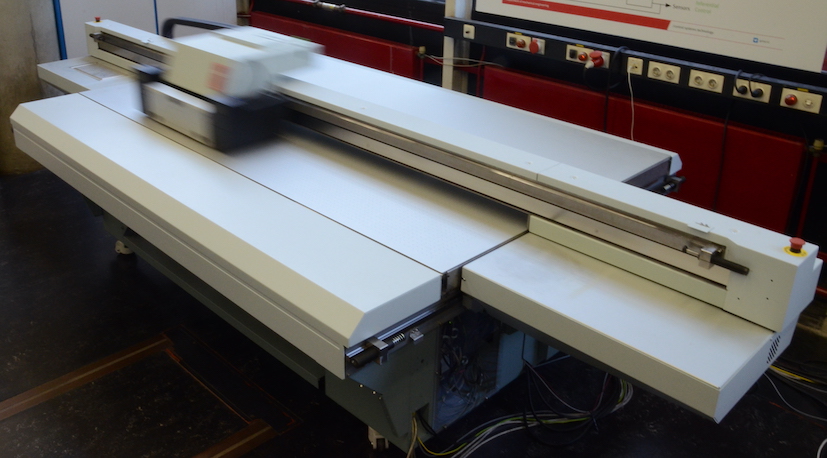}}
\caption{Multiple-input multiple-output industrial printer with varying setpoints.}
\label{fig:DSC_3417crop}
\end{figure}

\begin{figure}[t]
\centering
\includegraphics[width=.75\linewidth,page=1]{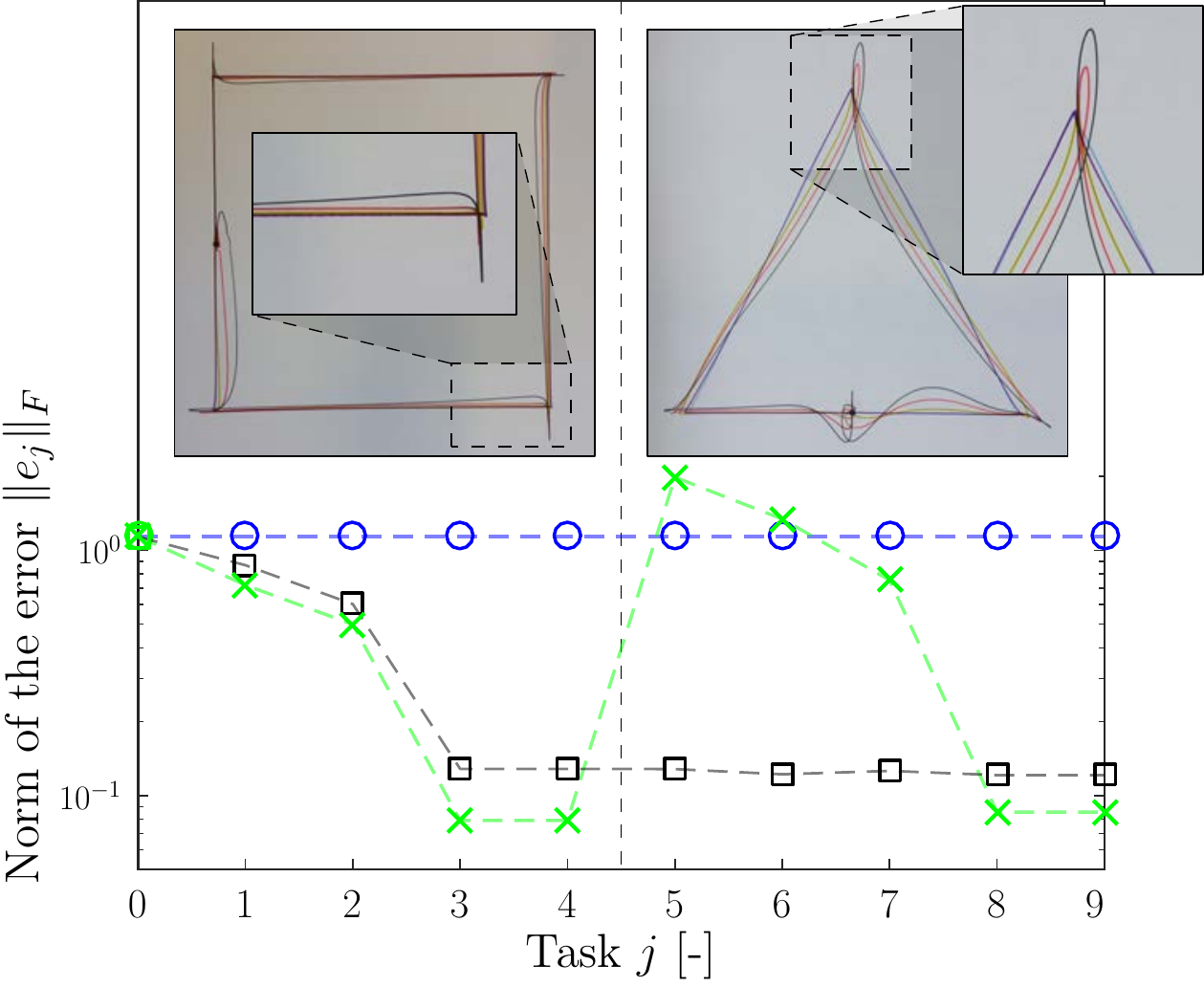}
\caption{Learning with varying references on the flatbed printer in \Fig~\ref{fig:DSC_3417crop}. In task $j = 0, 1, 2,3,4$, the setpoint is a square. Learning control (green crosses) substantially increases performance compared to feedback (blue circles). However, after a setpoint change to a triangle from task $j=5$ onwards, the ILC performance deteriorates and becomes worse compared to feedback only. ILC using basis functions (black squares) combines task flexibility and high performance.}
\label{fig:Figure11}
\end{figure}

Basis function ILC is applied to the multiple-input multiple-output industrial printer in \Fig~\ref{fig:DSC_3417crop}, see \Fig~\ref{fig:Figure11} for the results. It can be directly observed that a change of reference at task $j=5$ deteriorates the performance of the ILC approach of \Sec~\ref{sec:freqdomILC}, in fact the performance is worse compared to only using feedback. In sharp contrast, ILC with basis functions combines high performance and task flexiblility. 


\subsection{Multivariable learning}
\begin{figure}[t]
\centering
\includegraphics[width=.7\linewidth,page=1]{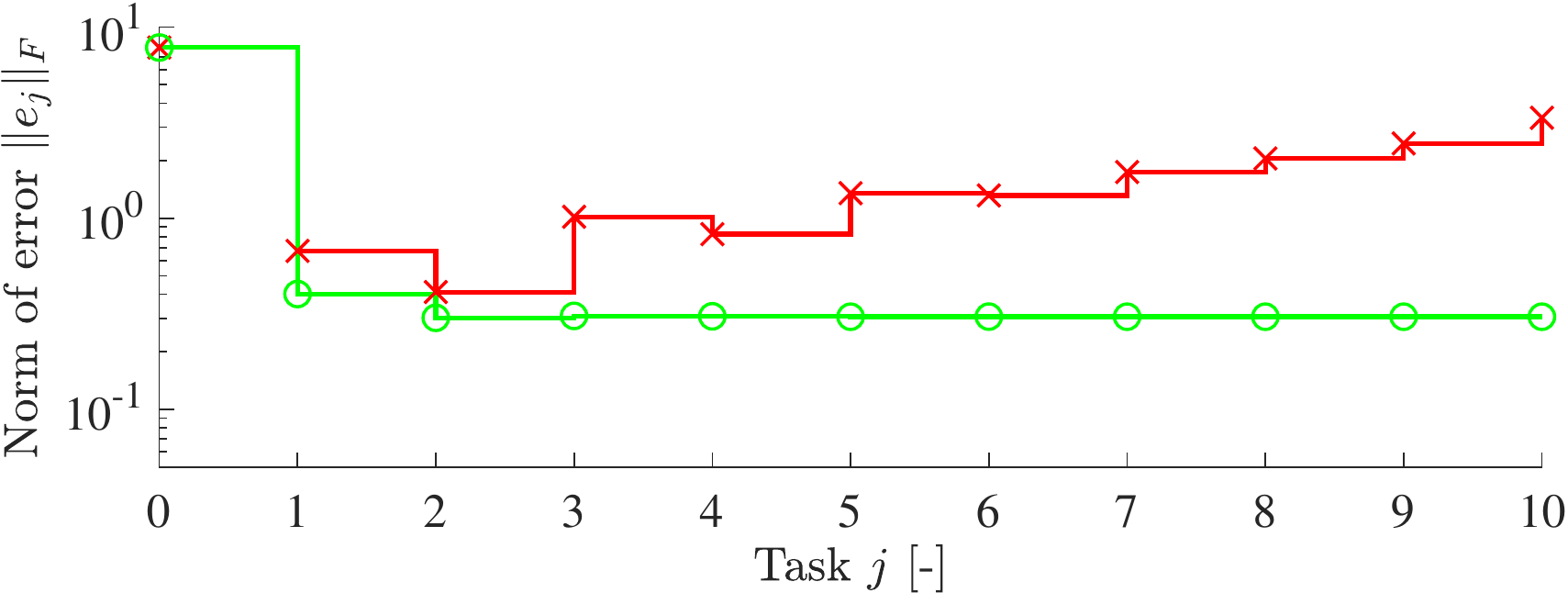}
\caption{Learning for multiple axes on the flatbed printer in \Fig~\ref{fig:DSC_3417crop}. Ignoring interaction by straightforward application of the approach in \Sec~\ref{sec:freqdomILC} leads to a divergent behavior (red crosses). Application of multivariable ILC leads to a converging error (green circles).}
\label{fig:ILC6}
\end{figure}

Industrial motion systems, including the printer in \Fig~\ref{fig:DSC_3417crop}, often have multiple actuators and sensors. Application of the scalar ILC approach of \Sec~\ref{sec:freqdomILC} to several input-output pairs sequentially or simultaneously may lead to disastrous results. Indeed, as is shown in \Sec~\ref{eq:ILCisfeedback}, ILC essentially is a feedback mechanism in the iteration domain. It is well-known that interaction is a key aspect in multivariable feedback control. In fact, since ILC is effective over a much larger bandwidth compared to traditional feedback control, see \Sec~\ref{sec:feedback}, interaction is substantially more important. 

In \cite{AAABlankenOom}, \cite{BlankenZunRozStrOom2019}, a multivariable ILC design framework is presented that extends the design philosophy of the approach in \Sec~\ref{sec:freqdomILC}. In particular, it uses nonparametric frequency response functions of the interaction elements to design $L$ and $Q$. 

Application of the approach of \Sec~\ref{sec:freqdomILC}  to the printer in  \Fig~\ref{fig:DSC_3417crop} leads to a diverging error, where $\|.\|_F$ denotes the Frobenius norm. In contrast, the approach in \cite{AAABlankenOom} leads to a converging error by accounting for interaction, see \Fig~\ref{fig:ILC6}.


\subsection{Inferential ILC}
Mechatronic systems often cannot be assumed rigid anymore due to extreme performance and accuracy requirements that are enabled by learning control. For instance, in the printer systems of \Fig~\ref{fig:printer} and \Fig~\ref{fig:DSC_3417crop}, the position is measured using an encoder, while performance is required at the printing location. This is referred to as  an inferential control situation \cite{OomenGraHen2015}. Also, sampled-data ILC is closely related, where only discrete observations of the performance variables are available \cite{OomenWijBos2009}.

ILC operates from one task to the next, also referred to as a batch-to-batch control approach, hence it can exploit measurements of the product. For instance, in printing systems, a scanner that immediately scans the print after each line \cite{BolderOomKoeSte2014c} can be used in conjunction with ILC. Although this is a very promising concept, a recent analysis \cite{BolderOom2016} has shown that traditional ILC approaches, including  the approach in \Sec~\ref{sec:freqdomILC}, lead to an internally unstable closed-loop situation if they are combined with feedback on the encoder. By adopting appropriate ILC structures, ILC can be directly applied to improve the performance of such inferential control problems. 


\subsection{Position-dependent and position-domain ILC}
Traditional ILC mainly involves linear and time-invariant systems, yet for many mechatronic systems this assumption is not satisfied. First, input-output dynamics may be nonlinear, including changing mass distributions in H-drive systems, such as the printer of \Fig~\ref{fig:DSC_3417crop} and positioning system in \cite{GrootwassinkWalSchBos2003}, as well as varying sensor locations such as in wafer scanners \cite{AAAVoorhoeveRozAanOom}. To address these nonlinear dynamics, both LPV ILC \cite{RozarioOomSte2017} and LTV ILC \cite{ZundertBolKoeOom2016b} are explored. Second, disturbances may be position-dependent, including errors induced by position-dependent commutation in motors and piezo-stepper actuators for long-range motion, which is investigated in ILC in \cite{AarnoudseStrVerOom2020}. In \cite{MoorenWitOom2020}, a closely related Gaussian process-based approach is developed. Third, measurements may be position-dependent as in encoder systems, which has been addressed through an intermittent sampling ILC approach \cite{StrijboschOom2019c}. Fourth, performance requirements may be position-dependent, e.g., in 3D printing, which has led to cross-coupled ILC \cite{BartonAll2008} and spatial ILC \cite{HoelzleBar2016}.




\subsection{Integrated data-driven modeling and learning}
The ILC design approach in \Sec~\ref{sec:freqdomILC} requires a parametric model, see in particular \Sec~\ref{sec:ILCdesign}. This allows for a robust design for a range of identified frequency response functions. However, it requires a parametric modeling step, which is relatively time consuming, and a robustness filter $Q$, which may lead to a residual asymptotic error and conservatism. 

To avoid the parametric modeling step, several approaches have been developed. In \cite{KimZou2013}, the learning filter $L$ is directly based on a nonparametric frequency response function, which is extended towards online estimation of the frequency response function in \cite{RozarioOom2019c}. A gradient-descent ILC approach is developed in \cite{BolderKleOom2018}, which estimates the gradient using dedicated experiments on the real system. In \cite{PootPorOom2020}, an actor-critic approach is developed for basis function ILC, see \Sec~\ref{sec:flextasks}, that avoids the use of explicit models by employing results from reinforcement learning, see, e.g., \cite{Recht2019} for related recent developments. 


%


\section{Conclusions}
Iterative learning control enables a major performance improvement for mechatronic systems. By employing inversion-based learning, fast convergence can be achieved within just a few iterations. Robustness can be enforced by employing frequency response function measurements for a large range of operating conditions or machine-to-machine variability. In recent years, ILC has been extended to facilitate widespread industrial deployment for mechatronic systems. These results are foreseen to enable a major performance improvement.

{\scriptsize
\emph{Acknowledgements}
The framework presented here is the result of a fruitful collaboration with many colleagues in academia and industry who are gratefully acknowledged. In particular, Leontine Aarnoudse, Frank Boeren, Lennart Blanken, Joost Bolder, Noud Mooren, Maurice Poot, Robin de Rozario, Nard Strijbosch, and Jurgen van Zundert have had a major role in the ideas outlined here. Financial support is provided by VIDI project number 15698, which is (partly) financed by the Netherlands Organisation for Scientific Research (NWO) and ECSEL-2016-1 under grant agreement number 737453 (I-MECH).}

\bibliographystyle{IEEETranS}


\begin{thebibliography}{10}
\providecommand{\url}[1]{#1}
\csname url@samestyle\endcsname
\providecommand{\newblock}{\relax}
\providecommand{\bibinfo}[2]{#2}
\providecommand{\BIBentrySTDinterwordspacing}{\spaceskip=0pt\relax}
\providecommand{\BIBentryALTinterwordstretchfactor}{4}
\providecommand{\BIBentryALTinterwordspacing}{\spaceskip=\fontdimen2\font plus
\BIBentryALTinterwordstretchfactor\fontdimen3\font minus
  \fontdimen4\font\relax}
\providecommand{\BIBforeignlanguage}[2]{{%
\expandafter\ifx\csname l@#1\endcsname\relax
\typeout{** WARNING: IEEEtranS.bst: No hyphenation pattern has been}%
\typeout{** loaded for the language `#1'. Using the pattern for}%
\typeout{** the default language instead.}%
\else
\language=\csname l@#1\endcsname
\fi
#2}}
\providecommand{\BIBdecl}{\relax}
\BIBdecl

\bibitem{AarnoudseStrVerOom2020}
L.~Aarnoudse, N.~Strijbosch, E.~Verschueren, and T.~Oomen, ``Commutation-angle
  iterative learning control for intermittent data: Enhancing piezo-stepper
  actuator waveforms,'' in \emph{\IFACWC[21]}, Berlin, Germany, 2020.

\bibitem{AbramovitchAndPaoSch2007}
D.~Abramovitch, S.~Andersson, L.~Pao, and G.~Schitter, ``A tutorial on the
  mechanics, dynamics, and control of atomic force microscopes,'' in
  \emph{\ACC[2007]}, New York, \NY, 2007, pp. 3488--3502.

\bibitem{ArimotoKawMiy1984}
S.~Arimoto, S.~Kawamura, and F.~Miyazaki, ``Bettering operation of robots by
  learning,'' \emph{\JRS}, vol.~1, no.~2, pp. 123--140, 1984.

\bibitem{BartonAll2008}
K.~L. Barton and A.~G. Alleyne, ``A cross-coupled iterative learning control
  design for precision mechatronics,'' \emph{\IEEECST}, vol.~16, no.~6, pp.
  1218--1231, 2008.

\bibitem{BienXu1998}
Z.~Bien and J.-X. Xu, \emph{Iterative Learning Control: Analysis, Design,
  Integration and Applications}.\hskip 1em plus 0.5em minus 0.4em\relax
  Norwell, \MA: Kluwer Academic Publishers, 1998.

\bibitem{BlankenBoeBruOom2017}
L.~Blanken, F.~Boeren, D.~Bruijnen, and T.~Oomen, ``Batch-to-batch rational
  feedforward control: from iterative learning to identification approaches,
  with application to a wafer stage,'' \emph{\IEEETM}, vol.~22, no.~2, pp.
  826--837, 2017.

\bibitem{BlankenKoeOom2020}
L.~Blanken, S.~Koekebakker, and T.~Oomen, ``Multivariable repetitive control:
  Decentralized designs with application to continuous media flow printing,''
  \emph{\IEEETM}, vol.~25, no.~1, pp. 294--304, 2020.

\bibitem{BlankenOom2020}
L.~Blanken and T.~Oomen, ``Kernel-based identification of non-causal systems
  with application to inverse model control,'' \emph{\AUT}, vol. 114, p.
  108830, 2020.

\bibitem{AAABlankenOom}
------, ``Multivariable iterative learning control design procedures: From
  decentralized to centralized, illustrated on an industrial printer,''
  \emph{\IEEECST}, To appear.

\bibitem{BlankenZunRozStrOom2019}
L.~Blanken, J.~van Zundert, R.~de~Rozario, N.~Strijbosch, and T.~Oomen,
  ``Multivariable iterative learning control: Analysis and designs for
  engineering applications,'' in \emph{Data-Driven Modeling, Filtering and
  Control}, C.~Novara and S.~Formentin, Eds.\hskip 1em plus 0.5em minus
  0.4em\relax {IET} The Institution of Engineering and Technology, 2019, pp.
  109--138.

\bibitem{BoerenBarKokOom2016}
F.~Boeren, A.~Bareja, T.~Kok, and T.~Oomen, ``Frequency-domain {ILC} approach
  for repeating and varying tasks: With application to semiconductor bonding
  equipment,'' \emph{\IEEETM}, vol.~21, no.~6, pp. 2716--2727, 2016.

\bibitem{BoerenBruDijOom2014}
F.~Boeren, D.~Bruijnen, N.~van Dijk, and T.~Oomen, ``Joint input shaping and
  feedforward for point-to-point motion: Automated tuning for an industrial
  nanopositioning system,'' \emph{\MECH}, vol.~24, no.~6, pp. 572--581, 2014.

\bibitem{BolderKleOom2018}
J.~Bolder, S.~Kleinendorst, and T.~Oomen, ``Data-driven multivariable {ILC}:
  Enhanced performance by eliminating ${L}$ and ${Q}$ filters,'' \emph{\IJRNC},
  vol.~28, no.~12, pp. 3728--3751, 2018.

\bibitem{BolderOom2015}
J.~Bolder and T.~Oomen, ``Rational basis functions in iterative learning
  control - with experimental verification on a motion system,''
  \emph{\IEEECST}, vol.~23, no.~2, pp. 722--729, 2015.

\bibitem{BolderOom2016}
------, ``Inferential iterative learning control: A {2D}-system approach,''
  \emph{\AUT}, vol.~71, pp. 247--253, 2016.

\bibitem{BolderOomKoeSte2014c}
J.~Bolder, T.~Oomen, S.~Koekebakker, and M.~Steinbuch, ``Using iterative
  learning control with basis functions to compensate medium deformation in a
  wide-format inkjet printer,'' \emph{\MECH}, vol.~24, no.~8, pp. 944--953,
  2014.

\bibitem{BristowThaAll2006}
D.~A. Bristow, M.~Tharayil, and A.~G. Alleyne, ``A survey of iterative learning
  control: A learning-based method for high-performance tracking control,''
  \emph{\CSM}, vol.~26, no.~3, pp. 96--114, 2006.

\bibitem{FruehPha2000}
J.~A. Frueh and M.~Q. Phan, ``Linear quadratic optimal learning control
  ({LQL}),'' \emph{\IJC}, vol.~73, no.~10, pp. 832--839, 2000.

\bibitem{Gawronski2004}
W.~K. Gawronski, \emph{Advanced Structural Dynamics and Active Control of
  Structures}.\hskip 1em plus 0.5em minus 0.4em\relax New York, \NY: Springer,
  2004.

\bibitem{GhazaeiKhoBer2017}
M.~Ghazaei, S.~Z. Khong, and B.~Bernhardsson, ``On the convergence of iterative
  learning control,'' \emph{\AUT}, vol.~78, pp. 266--273, 2017.

\bibitem{Gorinevsky2002}
D.~Gorinevsky, ``Loop shaping for iterative control of batch processes,''
  \emph{\CSM}, vol.~22, no.~6, pp. 55--65, 2002.

\bibitem{GrootwassinkWalSchBos2003}
M.~Groot~Wassink, M.~van~de Wal, C.~Scherer, and O.~Bosgra, ``{LPV} control for
  a wafer stage: Beyond the theoretical solution,'' \emph{\CEP}, vol.~13, pp.
  231--245, 2003.

\bibitem{GunnarssonNor2001}
S.~Gunnarsson and M.~Norrl{\"o}f, ``On the design of {ILC} algorithms using
  optimization,'' \emph{\AUT}, vol.~37, pp. 2011--2016, 2001.

\bibitem{GuoPetOomMis2018}
Y.~Guo, J.~Peters, T.~Oomen, and S.~Mishra, ``Control-oriented models for
  ink-jet {3D} printing,'' \emph{\MECH}, vol.~56, pp. 211--219, 2018.

\bibitem{Heertjes2016b}
M.~Heertjes, ``Data-based motion control of wafer scanners,'' in
  \emph{{IFAC-PapersOnLine} 49-13}, 2016, pp. 001--012.

\bibitem{HoelzleAllWag2011}
D.~J. Hoelzle, A.~G. Alleyne, and A.~J. Wagoner~Johnson, ``Basis task approach
  to iterative learning control with applications to micro-robotic
  deposition,'' \emph{\IEEECST}, vol.~19, no.~5, pp. 1138--1148, 2011.

\bibitem{HoelzleBar2016}
D.~J. Hoelzle and K.~L. Barton, ``On spatial iterative learning control via
  {2-D} convolution: Stability analysis and computational efficiency,''
  \emph{\IEEECST}, vol.~24, no.~4, pp. 1504--1512, 2016.

\bibitem{JanssensPipSwe2012}
P.~Janssens, G.~Pipeleers, and J.~Swevers, ``Initialization of {ILC} based on a
  previously learned trajectory,'' in \emph{\ACC[2012]}, Montreal, Canada,
  2012, pp. 610--614.

\bibitem{KimZou2013}
K.-S. Kim and Q.~Zou, ``A modeling-free inversion-based iterative feedforward
  control for precision output tracking of linear time-invariant systems,''
  \emph{\IEEETM}, vol.~18, no.~6, pp. 1767--1777, 2013.

\bibitem{LambrechtsBoeSte2005}
P.~Lambrechts, M.~Boerlage, and M.~Steinbuch, ``Trajectory planning and
  feedforward design for electromechanical motion systems,'' \emph{\CEP},
  vol.~13, pp. 145--157, 2005.

\bibitem{Longman2000}
R.~W. Longman, ``Iterative learning control and repetitive control for
  engineering practice,'' \emph{\IJC}, vol.~73, no.~10, pp. 930--954, 2000.

\bibitem{MaasMaasVooOom2017}
R.~{\SortAt{Maas}}van~der Maas, A.~van~der Maas, R.~Voorhoeve, and T.~Oomen,
  ``Accurate {FRF} identification of {LPV} systems: {nD-LPM} with application
  to a medical {X}-ray system,'' \emph{\IEEECST}, vol.~25, no.~4, pp.
  1724--1735, 2017.

\bibitem{MeulenTouBos2008}
S.~{\SortAt{Meulen}}van~der Meulen, R.~L. Tousain, and O.~H. Bosgra, ``Fixed
  structure feedforward controller design exploiting iterative trials:
  Application to a wafer stage and a desktop printer,'' \emph{\JDMC}, vol. 130,
  pp. 051\,006--1, 2008.

\bibitem{Moore1993}
K.~L. Moore, \emph{Iterative Learning Control for Deterministic Systems}.\hskip
  1em plus 0.5em minus 0.4em\relax London, \UK: Springer-Verlag, 1993.

\bibitem{MoorenWitOom2020}
N.~Mooren, G.~Witvoet, and T.~Oomen, ``{Gaussian} process repetitive control
  for suppressing spatial disturbances,'' in \emph{\IFACWC[21]}, Berlin,
  Germany, 2020.

\bibitem{MunnigschmidtSchEij2011}
R.~Munnig~Schmidt, G.~Schitter, and J.~van Eijk, \emph{The Design of High
  Performance Mechatronics}.\hskip 1em plus 0.5em minus 0.4em\relax Delft, The
  Netherlands: Delft University Press, 2011.

\bibitem{Oomen2018b}
T.~Oomen, ``Learning in machines,'' \emph{Mikroniek}, vol.~6, pp. 5--11, 2018.

\bibitem{Oomen2020}
------, ``Control for precision mechatronics,'' in \emph{Encyclopedia of
  Systems and Control}, 2nd~ed., J.~Baillieul and T.~Samad, Eds.\hskip 1em plus
  0.5em minus 0.4em\relax Springer Nature, 2020.

\bibitem{OomenGraHen2015}
T.~Oomen, E.~Grassens, and F.~Hendriks, ``Inferential motion control: An
  identification and robust control framework for unmeasured performance
  variables,'' \emph{\IEEECST}, vol.~23, no.~4, pp. 1602--1610, 2015.

\bibitem{OomenRoj2017}
T.~Oomen and C.~R. Rojas, ``Sparse iterative learning control with application
  to a wafer stage: Achieving performance, resource efficiency, and task
  flexibility,'' \emph{\MECH}, vol.~47, pp. 134--137, 2017.

\bibitem{OomenWijBos2009}
T.~Oomen, J.~{van de Wijdeven}, and O.~Bosgra, ``Suppressing intersample
  behavior in iterative learning control,'' \emph{\AUT}, vol.~45, no.~4, pp.
  981--988, 2009.

\bibitem{PaszkeRogGal2016}
W.~Paszke, E.~Rogers, and K.~Ga{\l}kowski, ``Experimentally verified
  generalized {KYP} lemma based iterative learning control design,''
  \emph{\CEP}, vol.~53, pp. 57--67, 2016.

\bibitem{PillonettoDinCheDenLju2014}
G.~Pillonetto, F.~Dinuzzo, T.~Chen, G.~De~Nicolao, and L.~Ljung, ``Kernel
  methods in system identification, machine learning and function estimation: A
  survey,'' \emph{\AUT}, vol.~50, no.~3, pp. 657--682, 2014.

\bibitem{PootPorOom2020}
M.~Poot, J.~Portegies, and T.~Oomen, ``On the role of models in learning
  control: Actor-critic iterative learning control,'' in \emph{\IFACWC[21]},
  Berlin, Germany, 2020.

\bibitem{Recht2019}
B.~Recht, ``A tour of reinforcement learning: The view from continuous
  control,'' \emph{Annual Review of Control, Robotics, and Autonomous Systems},
  vol.~2, pp. 253--279, 2019.

\bibitem{RogersGalOwe2007}
E.~Rogers, K.~Galkowski, and D.~H. Owens, \emph{Control Systems Theory and
  Applications for Linear Repetitive Processes}, ser. \LNCIS.\hskip 1em plus
  0.5em minus 0.4em\relax Berlin, Germany: Springer, 2007, no. 349.

\bibitem{RozarioOom2019c}
R.~{\SortAt{Rozario}}de~Rozario and T.~Oomen, ``Data-driven iterative
  inversion-based control: Achieving robustness through nonlinear learning,''
  \emph{\AUT}, vol. 107, pp. 342--352, 2019.

\bibitem{RozarioOomSte2017}
R.~{\SortAt{Rozario}}de~Rozario, T.~Oomen, and M.~Steinbuch, ``{ILC} and
  feedforward control for {LPV} systems: with application to a
  position-dependent motion system,'' in \emph{\ACC[2017]}, Seattle, \WA, 2017,
  pp. 3518--3523.

\bibitem{SilberSchSimAntHuaGueHubBakLaiBolCheLilHuiSifDriGraHas2017}
D.~Silver, J.~Schrittwieser, K.~Simonyan, I.~Antonoglou, A.~Huang, A.~Guez,
  T.~Hubert, L.~Baker, M.~Lai, A.~Bolton, Y.~Chen, T.~Lillicrap, F.~Hui,
  L.~Sifre, G.~van~den Driessche, T.~Graepel, and D.~Hassabis, ``Mastering the
  game of {Go} without human knowledge,'' \emph{Nature}, vol. 550, pp.
  354--359, 2017.

\bibitem{Stein2003}
G.~Stein, ``Respect the unstable,'' \emph{\CSM}, vol.~23, no.~4, pp. 12--25,
  2003.

\bibitem{SteinbuchMol2000}
M.~Steinbuch and R.~v.~d. Molengraft, ``Iterative learning control of
  industrial motion systems,'' in \emph{\MECHSYMP[1]}, Darmstadt, Germany,
  2000, pp. 967--972.

\bibitem{StrijboschOom2019c}
N.~Strijbosch and T.~Oomen, ``Intermittent sampling in iterative learning
  control: a monotonically-convergent gradient-descent approach with
  application to time stamping,'' in \emph{\CDC[58]}, Nice, France, 2019, pp.
  6542--6547.

\bibitem{TogaiYam1985}
M.~Togai and O.~Yamano, ``Analysis and design of an optimal learning control
  scheme for industrial robots: A discrete system approach,'' in
  \emph{\CDC[24]}, Fort Lauderdale, \FL, 1985, pp. 1399--1404.

\bibitem{AAAVoorhoeveRozAanOom}
R.~Voorhoeve, R.~de~Rozario, W.~Aangenent, and T.~Oomen, ``Identifying
  position-dependent mechanical systems: A modal approach with applications to
  wafer stage control,'' \emph{\IEEECST}, To appear.

\bibitem{WijdevenBos2010}
J.~{\SortAt{Wijdeven}}van~de Wijdeven and O.~Bosgra, ``Using basis functions in
  iterative learning control: Analysis and design theory,'' \emph{\IJC},
  vol.~83, no.~4, pp. 661--675, 2010.

\bibitem{ZundertBolKoeOom2016b}
J.~{\SortAt{Zundert}}van~Zundert, J.~Bolder, S.~Koekebakker, and T.~Oomen,
  ``Resource-efficient {ILC} for {LTI/LTV} systems through {LQ} tracking and
  stable inversion: Enabling large tasks on a position-dependent industrial
  printer,'' \emph{\MECH}, vol.~38, pp. 76--90, 2016.

\bibitem{ZundertOom2017e}
J.~{\SortAt{Zundert}}van~Zundert and T.~Oomen, ``On optimal feedforward and
  {ILC}: The role of feedback for optimal performance and inferential
  control,'' in \emph{\IFACWC[2017]}, Toulouse, France, 2017, pp. 6267--6272.

\bibitem{ZundertOom2018c}
------, ``On inversion-based approaches for feedforward and {ILC},''
  \emph{\MECH}, vol.~50, pp. 282--291, 2018.

\end{thebibliography}


\end{document}